\documentstyle[12pt,epsf]{article}
\advance\textheight by \topskip
\begin{document}

\rightline{UCSD/PTH-94-04}
\rightline{April 1994}

\vspace{.8cm}
\begin{center}
{\large\bf An Alternative Method of Extracting $V_{bu}$ \\
            from Semi-leptonic B Decay}

\vskip .9 cm

{\bf Jin Dai
\footnote{E-mail address: dai@higgs.ucsd.edu} }
 \vskip 0.1cm
Department of Physics 0319, 
University of California, San Diego \\
9500 Gilman Dr.
La Jolla, CA, 92093-0319

\end{center}

\vskip .6 cm
\centerline{\bf ABSTRACT}
\vspace{-0.7cm}
\begin{quotation}
We propose a new method of extracting $V_{bu}$ from measurements of
Semi-leptonic B decay which has much less theoretical uncertainties
than conventional methods. 
\end{quotation}

\normalsize
\newpage

The Kobayashi-Maskawa matrix element $V_{bu}$ is among the least
known standard model parameters. Knowing its exact value is very
important for understanding standard model CP violation and
radiative corrections. However, extracting $V_{bu}$ from 
experiment is difficult due to hadronic physics uncertainties. 

At the present time, $V_{bu}$ is measured in charmless   
semileptonic B decays, especially from the inculsive lepton spectrum. 
The advantage of using inclusive decay modes is that it can be
calculated using the parton model approach\cite{Geogi}, and 
nonperturbative effects can be systematically analysed using heavy
quark theory\cite{ManoWise}. However, in experiment, there is an 
overwhelming background from the decay process 
  \[  b \rightarrow c + l + \nu.  \] 
The only way to get a clean measurment is to concentrate on the
kinematic region where $b$ to $c$ decay is not allowed. In charmless B
decay, the lepton energy $E_l$ is kinematically allowed up to
$m_B/2$, and in $b$ to charm decay, $E_l$ is allowed up to $m_B -
m_D^2/2m_B$. That leaves available only about 330MeV of energy 
near the end point of lepton spectrum.

Unfortunately, it is the endpoint spectrum that is least known 
theoretically. In the parton model, the maximum $E_l$ allowed is $m_b/2$, 
where $m_b$ is the b quark mass instead of the B meson mass. For 
$m_b = 4.8 GeV$, this is 240MeV below the kinematic maximum. QCD radiative
corrections have been calculated\cite{radiative} and contain large logs 
near the parton model maximum. Beyond the parton model maximum, there is
no systematic way of calculating the lepton spectrum, and different
models lead to very different results\cite{models}. 

Instead of thinking of improving the theoretical estimate of the end
point lepton spectrum, we propose an alternative way to get a more
precise $V_{bu}$: Measure quantities which can be calculated
better theoretically. In this paper, we will assume that the neutrino 
momentum can be measured as well as the lepton momemtum in semileptonic
B decays. This is not unrealistic at electron-positron machines( e.g.
CLEO), where missing momentum and missing energy can be measured
and the neutrino mass shell condition can be used to reduce backgrounds
from particles leaking out of beam pipes.  We will 
show there is a better way to extract $V_{bu}$ which has few 
theoretical uncertainties.

First, with the neutrino momentum known, it is much easier
to separate the two quark processes $b\rightarrow cl\nu$
and $b\rightarrow ul\nu$ in semileptonic B decay 
\begin{equation}
       B \rightarrow l + \nu + X . 
\end{equation} 
The key observation is that if the underlying quark decay is
$b$ to $c$, then
\begin{equation}
       M_X^2 > m_D^2 . 
\end{equation}
For $b$ to $u$ decay, at the quark level, the process is dominated
by one outgoing $u$ quark or one $u$ quark and one gluon going almost
parallel. In either case,
\begin{equation}
       M_X^2 \approx 0 .
\end{equation} 
When the momenta of the B meson, the lepton and the neutrino are
all known, the 4-momentum of X is determined. Therefore we propose
the following cut which corresponds to excluding the kinematic region
defined by equation (2):
\begin{equation}
     E_l + E_\nu > \frac{m_B^2 - m_D^2 
         + 2E_l E_\nu (1-\cos\theta)}{2m_B} .
\end{equation}
where $E_l$ and $E_\nu$ are lepton and neutrino energies in the
rest frame of the B meson and $\theta$ is the angle between the
directions of the lepton and the neutrino.
This cut will exclude the $b$ to $c$ decay background and keep most
of the $b$ to $u$ decay events.   

Based on the matrix elements by Ali and Pietarinen\cite{radiative},
we did a numerical calculation of lepton and neutrino distribution
including QCD radiative corrections. It turns out that about 95\% of 
the $b\rightarrow u$ events will survive the cut mentioned above.   
Without knowing $\theta$, you can still exclude $b$ to $c$ decay 
kinematically by using the cut obtained by setting 
$\cos\theta = -1$ in equation (4),  
but only around 40\% of events are left. Without any
knowlede of the neutrino, as is the way that current experiments
are done, one can use only a small portion of the events.( For 
reasons explained above, theory can not predict the exact amount 
of events near the lepton end point.)
Background from charm quark leptonic decay can be excluded by cutting
on a high invariant mass of the the lepton-neutrino pair, if
it turns out to be statistically favorable. 

The parton model differential cross section 
$ d\sigma/dE_l dE_\nu  $  has large logs when
$E_l$ or $E_\nu$ approach $m_b/2$, but now we don't have
to focus on the edge of the kinematically allowed region where the theory
has large uncertainties. Instead, we can make the cut
\begin{equation}
         E_l, E_\nu < \frac{m_B}{2} - \delta
\end{equation}
where $\delta$ should be a few hundred MeV. We will get a partial
decay rate which is almost free of theoretical uncertainty, and
be able to compare to experiment to extract $V_{bu}$! The remaining
uncertainty comes from uncertainty in $m_b$, which may be
determind by studying the shape of the $b$ to $c$ semi-leptonic
decay spectrum. 

It is up to experimental physicists to decide whether the neutrino
momentum can be reconstructed with a reasonable efficiency, but if
it can be done, we can have a much better measurement of $V_{bu}$.
Having good detector coverage will be important on future B factories
for using this method.

After this work was completed, we received \cite{BZ}, in which
the authors suggest measuring the total final state
hadron energy. This corresponds to making the following cut
\[     E_l + E_\nu > m_B - m_D     \]
in which 25\% of the events (accordding to our calcultion) near the 
end point will survive. Of course, if one can identify each hadronic
track of the B decay, one can measure $m_X^2$ directly.

\newpage
\begin{figure}[htb]
\centerline{\epsfysize = 16cm,     \epsfbox{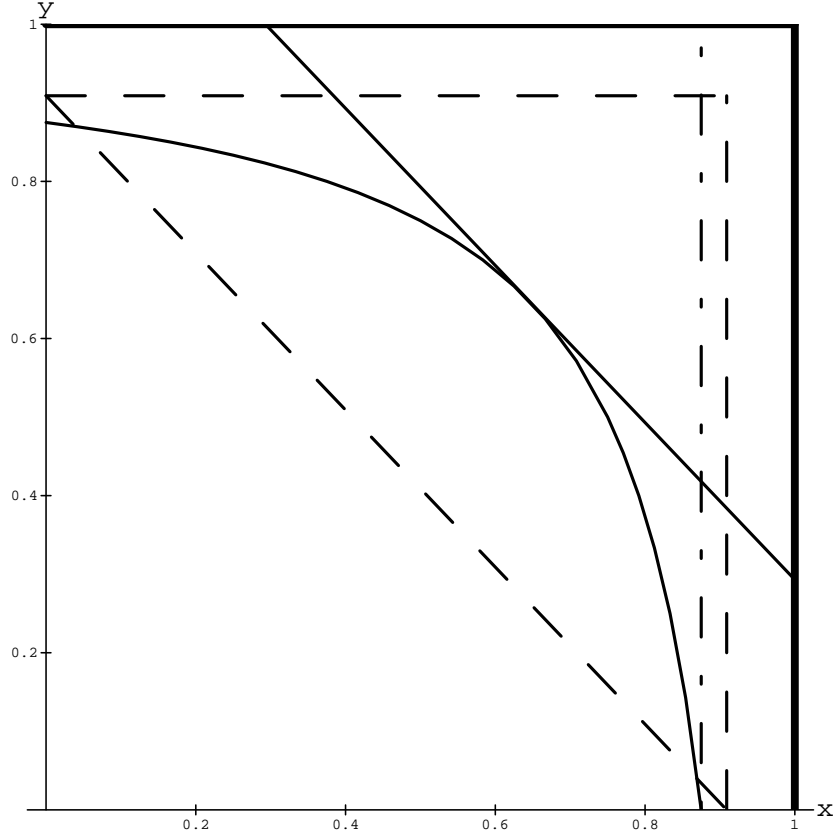}}
\caption{The phase space plot: $x = 2E_l/m_B$, $y = 2E_\nu /m_B$.
         Lepton and neutrino energy are kinematically  allowed
         inside the square, however, leading order parton model
         only allows final states inside the dashed triangle,
         therefore that is where most of the events are expected 
         to be. The conventional method of extracting $V_{bu}$
         utilizes the narrow region to the right of the dot-dashed
         line. Cutting on the total final state hadron energy
         makes use the region above the solid straight line
         while still excludes $b \rightarrow c$ decay. 
         Knowing both the lepton and neutrino energy but not
         the angle between them will allow us to use events
         above the solid curved line. Below the solid curved
         line, one can still use the angle to distinguish the
         two types of decay. }
\end{figure}  
         
\section*{Acknowledgements}
The Author would like to thank A. Manohar for extensive discussion,
H. Paar for discussion of CLEO experiment, and M. Luke for
useful conversation. Work was supported by DOE under grant 
DE-FG03-90ER40546.
\vskip 1cm



\begin{thebibliography}{100}
\bibitem{Geogi} J. Chay, H. Georgi and B. Grinstein, Phys. Lett. 
                B247, 399(1990). 

\bibitem{ManoWise} I.I. Bigi, M. Shifman, N.G. Uraltsev and
                   A.I. Vainshtein, Phys. Rev. Lett. 71,
                   496(1993). \\
                   A. Manohar and M.B. Wise, Phys. Rev D49,
                   1310(1994). 

\bibitem{radiative} A. Ali and E. Pietarinen, Nucl. Phys. B154,
                    519(1979). \\
                    N. Cabibbo, G. Corbo and L. Maiani, Nucl. Phys.
                    B155, 93(1979); G. Corbo, ibid. B212, 99(1983)\\
                    M. Jezabek and J.H. Kuhn, Nucl. Phys. B320, 
                    20(1989). 

\bibitem{models} B. Grinstein, N. Isgur and M.B. Wise, Phys. Rev. 
                 lett. 56, 258(1986); 
                 N. Isgur, D. Scora, B. Grinstein and M.B. Wise,
                 Phys. Rev. D39, 799(1989).\\
                 G. Altarelli, N. Cabibbo, G. Corbo, L. Maiani and
                 G. Martinelli, Nucl. Phys. B208.   

\bibitem{BZ}   A. Bouzas and D. Zappala,  UCLA/94/TEP/10

\end{thebibliography}
\end{document}